\documentstyle[prd,aps,preprint]{revtex}
        		
\newcommand\ee{\end{equation}}
\newcommand\be{\begin{equation}}
\newcommand\eea{\end{eqnarray}}
\newcommand\bea{\begin{eqnarray}}

\newcommand\km{\,\mbox{km}}

\newcommand\GeV{\,\mbox{GeV}}

\newcommand\Mpl{M_{\rm Pl}}

\newcommand\lsim{\mathrel{\rlap{\lower4pt\hbox{\hskip1pt$\sim$}}
    \raise1pt\hbox{$<$}}}
\newcommand\gsim{\mathrel{\rlap{\lower4pt\hbox{\hskip1pt$\sim$}}
    \raise1pt\hbox{$>$}}}

\def\dslash{\not{\hbox{\kern-2pt $\partial$}}}
\def\Dslash{\not{\hbox{\kern-4pt $D$}}}
\def\Oslash{\not{\hbox{\kern-4pt $O$}}}
\def\Qslash{\not{\hbox{\kern-4pt $Q$}}}
\def\pslash{\not{\hbox{\kern-2.3pt $p$}}}
\def\kslash{\not{\hbox{\kern-2.3pt $k$}}}
\def\qslash{\not{\hbox{\kern-2.3pt $q$}}}
 \newtoks\slashfraction
 \slashfraction={.13}
 \def\slash#1{\setbox0\hbox{$ #1 $}
 \setbox0\hbox to \the\slashfraction\wd0{\hss \box0}/\box0 }
 
\def\eeq{\end{equation}}
\def\beq{\begin{equation}}

\begin{document}

\preprint{FERMILAB-PUB-97/234-A}
\draft
\tighten

\title{Inflation and the Nature of Supersymmetry Breaking}
\author{Antonio Riotto\footnote{Email: riotto@fnas01.fnal.gov.}}
\address{NASA/Fermilab Astrophysics Center,\\ Fermilab
National Accelerator Laboratory, 
Batavia, Illinois~~60510.
}
\date{July 1997}
\maketitle

\begin{abstract}
The scale at which supersymmetry is broken   and the mechanism by which supersymmetry breaking is fed down to the observable sector has rich implications on   the way Nature may have  chosen to accomplish inflation.
We discuss  a simple model for slow rollover inflation which is minimal in the sense that the inflaton may be  identified with the  field responsible for the generation of the $\mu$-term. Inflation takes place at very late times and is characterized by a very low reheating temperature.  This property is crucial to solve the gravitino problem and may  help to  ameliorate the cosmological moduli problem. The COBE normalized value of the vacuum energy  driving inflation is naturally of the order of $10^{11}$ GeV. This favors the $N=1$ supergravity scenario where supersymmetry breaking is mediated by gravitational interactions. Nonetheless, smaller values of the vacuum energy are not  excluded by present data on the temperature anisotropy and the inflationary scenario may be implemented in the context of new recent ideas about gauge mediation where the standard model gauge interactions can serve as the messangers of supersymmetry breaking. In this class of models supersymmetry  breaking masses are usually propor
tional to the $F$-term of a gauge singlet superfield. The same  $F$-term  may provide the vacuum energy density necessary to drive inflation. The spectrum of density perturbations is characterized by a spectral index which is significantly displaced from one.  The measurements of the temperature anisotropies in the cosmic microwave background radiation at the accuracy expected to result from the  planned missions will be able to confirm or disprove this prediction  and to help in getting   some deeper  insight into the nature of   supersymmetry breaking.

\end{abstract}
\newpage
\baselineskip=18pt
\section{Introduction}
It is widely accepted that the structure of the 
 standard model of gauge interactions  is not complete.  Only to mention a few drawbacks, 
the theory has a plenty of unknown parameters, it does not describe the origin of fermion masses and why the number of generations is three.  The spontaneous symmetry breaking is triggered by a   light fundamental
scalar, the Higgs field, which is  something difficult to reconcile with our current
understanding of field theory.  Finally, gravity is not   incorporated.  It is tempting to speculate that a new (but   yet
undiscovered) symmetry,   supersymmetry (SUSY)   \cite{susy}, may provide answers to these fundamental questions.   Supersymmetry
is the only framework in which 
 light fundamental scalars appear natural. It addresses the question
of parameters:   first, unification of gauge couplings
works much better with than without supersymmetry;
second,
it is easier to attack questions such as
fermion masses in supersymmetric theories, in
part simply due to the presence of fundamental
scalars.  Supersymmetry seems to
be intimately connected with gravity.  
So there are a number of arguments that suggest
that nature might be supersymmetric, and that supersymmetry
might manifest itself at energies of order the weak
interaction scale.  

Another fundamental question is whether  supersymmetry  plays a fundamental role
at the early stages of the    evolution of the universe and, more specifically, during inflation \cite{abook}. 
The vacuum energy driving inflation is generated by
a scalar field $\phi$  displaced from the minimum of a 
potential $V\left(\phi\right)$. Quantum fluctuations of the inflaton field imprint a
nearly scale invariant spectrum  of fluctuations on the background space-time
metric. These fluctuations may be responsible for the generation of structure
formation. However, the level of density and temperature fluctuations observed
in the present universe, $\delta_H=1.94\times  10^{-5}$, require the inflaton
potential to be extremely flat \cite{lr}.
 For instance, in the chaotic inflationary scenario \cite{chaotic} where the
inflaton potential is $V=\lambda\phi^4$ and the scalar field sits initially at
scales of order of the Planck scale, the dimensionless self-coupling $\lambda$
must be of order of $10^{-13}$ to be consistent with observations.
 The inflaton field must be coupled to other fields in order to ensure the
conversion of the vacuum energy into radiation at the end of inflation, but
these couplings must be very small, otherwise loop corrections to the inflaton
potential spoil its flatness. This is where supersymmetry comes to rescue. 

 While the necessity of introducing very small
parameters to ensure the extreme flatness of the inflaton potential seems very
unnatural and fine-tuned in  most non-supersymmetric theories, this technical
naturalness may be achieved in supersymmetric models. Indeed,  the
nonrenormalization theorem guarantees that a fundamental object in supersymmetric theories, the  superpotential,  is not
renormalized to all orders of perturbation theory \cite{grisaru79}. In other words, the nonrenormalization theorems in exact global supersymmetry guarantee that we can fine-tune any parameter at the tree-level and this fine-tuning will not be destabilized by radiative corrections at any order in perturbation theory. Therefore, inflation in the context of supersymmetric theories seems, at least technically speaking, more natural than in the context of non-supersymmetric theories. 

On the other hand we know that the world is not globally supersymmetric: experimental searches show that scalar partners of the known fermions must be heavier than about 100 GeV. In Nature, supersymmetry must be a broken symmetry. 
Moreover, supersymmetry is necessarily broken during the inflationary stage by the vacuum energy density driving inflation. It seems  therefore  evident  that the  way and the scale at which supersymmetry are crucial ingredients  in any inflationary scenario.   

 The most  common and popular  approach is
to implement supersymmetry breaking in some  hidden
sector where some $F$-term gets a vacuum expectation value (VEV)  and then transmit it to the standard model sector  by
gravitational interactions. This is the so-called hidden $N=1$ supergravity scenario \cite{susy}. If one arranges the parameters in the hidden sector in such a way that the typical $\langle F\rangle $-term is of the order of $\langle F\rangle^{1/2} \sim\sqrt{(1\:{\rm TeV}) \Mpl}\sim 10^{11}$ GeV, where $\Mpl=2.4\times 10^{18}$ GeV,   the gravitino mass $m_{3/2}$ turns out to be of the order of  the TeV scale. 

An alternative to the hidden $N=1$ supergravity scenario is to break supersymmetry dynamically at lower energies.   
It has been known for a long time that supersymmetry breaking in four dimensions may be dynamical \cite{ads,reviewdsb}. There are many ways dynamical supersymmetry may arise. In theories like string theory, the potential is characterized by many classicaly flat directions. Typically, the potentials generated along these flat directions fall down to zero at large values of the fields. Perhaps, the most familiar example of this kind is the dilaton of string theory whose potential goes to zero in the weak coupling regime \cite{weak}. These potentials, however, must be stabilized by some mechanism and so far no compelling model has been proposed. 
Alternatively, models are known in which supersymmetry is broken without flat directions \cite{ads} and no need of complicated stabilization mechanisms is asked for. One can think of breaking supersymmetry 
at low energies. In such a case, gauge interactions can serve as the messangers of supersymmetry breaking 
and the SUSY breaking mediators are fields that transform nontrivially under
the  standard-model gauge group \cite{gmsb}. These are the so-called gauge mediated supersymmetry breaking (GMSB) models. The cosmology of   this latter class
of models has only been partially explored. In  particular,  a simple model for inflation in the context of GMSB models has been recently sketched in \cite{dr}. 

It is the purpose of this paper  to analyzed in more details and to extend the results of ref. \cite{dr} and to show that it is possible to construct an inflationary model    
  satisfying a sort of  ``minimal principle" which  requires   that the inflaton field
should not be an extra degree of freedom inserted   in some supersymmetric theory of particle physics just to drive inflation \cite{dr}:    the inflaton field may be identified with the same scalar responsible for the generation of the $\mu$-term present in the effective superpotential of the minimal supersymmetric standard model, $ W_\mu= \mu H_U H_D$. 
This opens  the exciting  possibility of connecting  a theory which could be tested at accelerators
with measurements of the  temperature anisotropy in the
cosmic microwave background and related measurements of the two-point correlation function. 

Quantum fluctuations of the inflaton field give rise to  temperature perturbations in the CMB at the level of  $\delta_H=1.94\times  10^{-5}$.   It is very intriguing  that the correct  level of density perturbations   is predicted if  the energy  $V^{1/4}$ driving  inflation is of the order of $10^{11}$ GeV, the same scale  springing out in the  $N=1$ supergravity scenario where supersymmetry breaking is mediated by gravitational interactions.  Present experimental data, however, do not exclude
smaller values of $V^{1/4}$, of the order of  $ 10^{10}$ GeV. We will argue that    this relatively high scale is not  unnatural in the context of GMSB models and that, if supersymmetry breaking is mediated by gauge interactions, 
the  energy density driving inflation may be identified  with the  same   $F$-term responsible for the spectrum of the superpartners in the low-energy effective theory. 

 The fluctuations arising from the quantum fluctuations of the inflaton field may be characterized by a power spectrum, which is the Fourier transform of the two-point density autocorrelation function. The power spectrum $\delta_H^2$  has the primordial form proportional to  $k^{(n-1)}$,  where $k$ is the amplitude of the Fourier wavevector and $n$ denotes the spectral index. 
We will show that the spectral index may be significantly smaller than 1.  This means that    the inflationary stage generates density perturbation which are far from being scale invariant. In the  model discussed in this paper, larger values of $|n-1|$ are associated to smaller values of the SUSY breaking scale. This prediction has the advantage that 
the  measurements of the temperature anisotropies in the cosmic microwave background (CMB) at the accuracy expected to result from two planned missions, the Microwave Anisotropy Probe (MAP) and PLANCK (formerly COBRAS/SAMBA), will allow us to confirm or disprove the model and, if confirmed,   to help in getting   some deeper  insight into the nature of   supersymmetry breaking.

Another distinguishing property of the inflationary scenario investigated  in this paper is   that inflation takes place at late times, when the Hubble parameter $H$ is $10^2$  GeV or so. This late period of inflation has dramatic effects
on the gravitino problem which is a  common  conundrum in both the $N=1$ supergravity scenario and in the GMSB models.  

In GMSB model building it is usually assumed that     the gravitino is    very light and certainly the lightest supersymmetric particle. This happens because the gravitino mass is given by   $m_{3/2}\sim F/\Mpl$  and the scale of supersymmetry breaking is taken to be of the order of  $\sqrt{F}\sim 10^6$ GeV.  The resulting gravitino mass is in the KeV range. However, if one   is willing to identify the scale of inflation with the 
same $F$-term responsible for the spectrum of the superpartners,   the COBE normalization predicts  values for the $F$-term higher than usually assumed. 
As we will show, this does not  exclude the possibility of implementing inflation in the framework of  GMSB theories.   The gravitino  mass
may be   of the order of 50 GeV. It is remarkable that such heavy gravitinos do not pose any cosmological problem. 
If the gravitino is  cosmologically stable and it  is thermalized in the early universe and not diluted by any mechanism, its mass density may exceed the closure limit $\Omega_{3/2}\lsim 1$. Since the number density of gravitinos is fixed once they  are thermalized, the above argument sets a stringent  upper bound on the gravitino mass,  $m_{3/2}\lsim 2$ keV if no dilution is present \cite{pargel}. 
However, we will show that such stable gravitinos are efficiently diluted during the inflationary stage  and they are not produced in the subsequent stage of reheating. This happens because   the reheating temperature turns out to be very low, of the order of 1 GeV.  

If gravitinos are cosmologically unstable,  which is certainly the case in  the 
$N=1$ supergravity scenario and is  a possibility  in  GMSB models,   
their decays do not   jeopardize the nice predictions of big-bang nucleosynthesis being the reheating temperature very small. 

We will also  address shortly  the cosmological problem represented by moduli ${\cal M}$.
In models of gauge mediation, if we assume that the underlying theory is a string theory,
the cosmological moduli problem is even more  than in the usual supergravity scenarios \cite{moduli}. The  period of inflation   may take place
at a sufficiently   late stage of the universe,  $H\lsim m_{{\cal M}}$,  that
the number density of string moduli is  exponentially reduced during inflation and by  the subsequent entropy production at the reheat stage \cite{randall}. However, this may be  not enough, since the minimum
of the moduli potential, generically, will be shifted by an amount of order $\Mpl$ during
inflation and
it is probably necessary to invoke symmetry reasons
so that the minima  during inflation and at the present day coincide  to a high degree of accuracy.

Despite the  low value of the reheating temperature,  the  production of the baryon asymmetry may occur during the stage at which the universe is reheated up and the standard big-bang era begins. We will show in details that this is possible if  the inflaton field has  nonrenormalizable which contain  $CP$-violation and baryon number violation.  

The paper is organized as follows. In section II we shortly describe  the main  
features of the $N=1$ supergravity scenario and  of GMSB models; the $\mu$-problem and one possible solution to it are addressed in section III; sections IV and V  deal with the inflationary stage and the post-inflationary stage, respectively; conclusions and outlook are presented in section VI. 

\section{The scale of supersymmetry breaking}

As we already mentioned,  the  scale of supersymmetry
breaking, is usually much larger than the weak scale. 

 The most convenient and widely used  approach is
to implement supersymmetry breaking in some  hidden
sector and then transmit it to the standard model sector  by
gravitational interactions. The scale of breaking is  of order
$10^{11}$ GeV.  An alternative possibility which has attracted so much attention recently  is that supersymmetry breaking is transmitted
via the gauge interactions of a distinct messenger sector
which contains fields that transform nontrivially under
the  standard-model gauge group. Let us now shortly review these two different scenarios. 

\subsection{Hidden $N=1$ supergravity scenario}

When supersymmetry is promoted from a global to a local symmetry, gravity is automatically taken into account and, accordingly, the theory is dubbed  supergravity. The supergravity lagrangian is defined in terms of the K\"{a}hler potential $K\left(\Phi,\Phi^{\dagger}\right)$ which can be split according to
\begin{equation}
K\left(\Phi,\Phi^{\dagger}\right)=d\left(\Phi,\Phi^{\dagger}\right)
+{\rm ln}|W(\Phi)|^2,
\end{equation}
where $\Phi^A\equiv (\phi^i,y^a)$ are the left-handed chiral superfields of the hidden $(\phi^i)$ and observable $(y^a)$ sectors.  Here $d$ and $W$ (the superpotential)
must be chosen to be invariant under the symmetries of the theory. Notice that the  K\"{a}hler potential in the hidden sector needs not to be minimal, $K_{{\rm min}}=\sum_A \phi_A^\dagger\phi_A$. 
Since higher order terms (suppressed by powers of $M_{{\rm Pl}}$) are not forbidden by any symmetry (or in any case they are expected to be generated by radiative corrections), they will surely be present at some level. 

From $K$ one can derive the scalar  potential $V$
\begin{equation}
\label{sugra}
V={\rm exp}\left(d/\Mpl^2\right)\left[F^{A\dagger} (d^{-1})_A^B F_B-3 \frac{|W|^2}{\Mpl^2}\right]+{\rm D-terms},
\end{equation}
where
\begin{eqnarray}
F_A&=&\frac{\partial W}{\partial\Phi^A}+\left(\frac{\partial d}{\partial\Phi^A}\right)\frac{W}{\Mpl^2},\nonumber\\
(d^{-1})_A^B&=& \left(\frac{\partial^2 d}{\partial\Phi^A\partial\Phi_A^\dagger}\right)^{-1}.
\end{eqnarray}

Requiring that the low-energy lagrangian for the matter fields is not multiplied by powers of $\Mpl$ defines the dependence on $\Mpl$ of $W$ and $d$  \cite{soni} 
\begin{eqnarray}
W(\xi,y)&=&\Mpl^2 W^{(2)}(\xi)+\Mpl W^{(1)}(\xi)+ W^{(0)}(\xi,y),\nonumber\\
d(\xi,\xi^\dagger,y,y^\dagger)&=& \Mpl^2 d^{(2)}(\xi,\xi^\dagger) +\Mpl d^{(1)}(\xi,\xi^\dagger) + d^{(0)}(\xi,\xi^\dagger,y,y^\dagger),
\end{eqnarray}
where $\xi^i\equiv \phi^i/\Mpl$. In addition, to obtain a renormalizable low-energy theory, we must require kinetic terms for the $y^a$ fields of the form \cite{soni}
\begin{equation}
 d^{(0)}(\xi,\xi^\dagger,y,y^\dagger)=y^a\Lambda_a^b(\xi,\xi^\dagger)y_b^\dagger+\left(\Gamma(\xi,\xi^\dagger,y)+{\rm h.c.}\right),
\end{equation}
with the vacuum expectation value $\langle \Lambda_a^b\rangle=\delta_a^b$. Finally, the $\phi^i$ fields being gauge singlets, gauge invariance requires
$\Lambda_a^b$ to be diagonal. If there are no mass scales in the theory other than $\Mpl$ and those induced by some spontaneous symmetry breaking (this is what happens in superstring-inspired theories), the renormalizable self couplings of the 
light fields $y^a$ is of the form \cite{gm}
\begin{eqnarray}
W^{(0)}(\xi,y)&=&\sum_n\: c_n(\xi) g_n^{(3)}(y),\nonumber\\
\Gamma(\xi,\xi^\dagger,y)&=&\sum_m\: c_m^\prime(\xi,\xi^\dagger)g_m^{(2)}(y),
\end{eqnarray}
where $g_n^{(3)}(y)$ and $g_m^{(2)}(y)$ are, respectively, the trilinear and bilinear terms in $y^a$ allowed by the symmetries of the theory. From these expressions one can show that,  in the limit $\Mpl\rightarrow\infty$ and after the hidden sector gauge singlets have acquired a vacuum expactation value such that $\langle\xi_i\rangle\sim 1$,
 the soft SUSY breaking terms in the effective potential of the light fields of the ordinary matter sector are characterized by  a common scale $m_{3/2}$, the gravitino mass
\begin{equation}
 m_{3/2} \equiv \langle{\rm e}^{d^{(2)}/2} W^{(2)}\rangle.  
\end{equation}
If one chooses the parameters in the hidden sector in such a way that the typical $\langle F\rangle$-term is of the order of $\langle F\rangle^{1/2}\sim 10^{11}$ GeV,  the gravitino mass  turns out to be of the order of $m_{3/2} \equiv \langle{\rm e}^{d^{(2)}/2} W^{(2)}\rangle\simeq 10^3$ TeV.

\subsection{Low energy dynamical supersymmetry breaking and gauge mediation}

An alternative approach  to the supergravity
approach is to suppose that supersymmetry is broken
at some low energy, with gauge interactions
serving as the messengers of supersymmetry
breaking \cite{gmsb}. The basic
idea is  that
the theory contains  new fields and interactions which break
supersymmetry.  Some of these fields are taken
to carry ordinary
standard model quantum numbers, so that
 ordinary squarks, sleptons and gauginos
can couple to them through gauge loops.  

The minimal gauge mediated supersymmetry breaking  models are defined by three sectors: {\it (i)} a secluded
sector that breaks supersymmetry; {\it (ii)} a messenger sector that serves
to communicate the SUSY breaking to the standard model and {\it (iii)} the
 standard model sector.
The minimal messenger sector consists of a single
${\bf 5}+\bar{{\bf 5}}$ of $SU(5)$
(to preserve gauge coupling constant unification),
{\it i.e.} color triplets, $q$ and $\bar{q}$,
and weak doublets $\ell$ and $\bar{\ell}$
with their interactions with a singlet superfield $X$ determined by the following superpotential:
\begin{equation}
W=\lambda_1 X\bar{q}q+\lambda_2 X\bar{\ell}\ell.
\end{equation}
When the field $X$ acquires a vacuum expectation value for both its scalar and auxiliary components,
$\langle X\rangle$ and $\langle F_X\rangle$ respectively,
the spectrum for $(q,\ell)$ is rendered non-supersymmetric.
Integrating out the messenger sector gives rise to gaugino
masses at one loop and scalar masses at two loops. For gauginos, we have
\begin{equation}
M_j(\Lambda)=k_j\frac{\alpha_j(\Lambda)}{4\pi}\Lambda,\:\:\:j=1,2,3,
\end{equation}
where $\Lambda= \langle F_X\rangle/\langle X\rangle$,
$k_1=5/3$, $k_2=k_3=1$ and $\alpha_1=\alpha/\cos^2\theta_W$. For the scalar
masses one has
\begin{equation}
\widetilde{m}^2(\Lambda)=2 \sum_{j=1}^3\:C_j
k_j\left[\frac{\alpha_j(\Lambda)}{4\pi}
\right]^2\Lambda^2,
\end{equation}
where $C_3=4/3$ for color triplets, $C_2=3/4$ for weak doublets
(and equal to zero otherwise) and $C_1=Y^2$ with $Y=Q-T_3$.

Because the scalar masses are functions of only the gauge quantum numbers,
 the flavour-changing-neutral-current processes are naturally suppressed in agreement with
experimental bounds.
The reason for this suppression is that the gauge interactions
induce flavour-symmetric supersymmetry-breaking terms in the observable sector at
$\Lambda$ and, because this scale is small, only a slight asymmetry
is introduced by renormalization group extrapolation to low energies.
This is in contrast to the supergravity scenarios
where one generically needs to invoke additional flavor symmetries to
achieve the same goal.
  
If squark and gauginos
 have to be around 1 TeV, the scale $\Lambda$ should be of the order of $10^3$
TeV.

It is important to notice that  this does not necessarily mean that $\sqrt{\langle F_X\rangle}$ and $\langle X\rangle$   must be of the same order of magnitude of $\Lambda$, being only their ratio fixed to be around $10^3$ TeV: the hierarchy $\sqrt{\langle F_X\rangle},\langle X\rangle\gg\Lambda$ is certainly allowed \cite{raby}. Large values of $\sqrt{\langle F_X\rangle}$ and $\langle X\rangle\gg\Lambda$ may be obtained if, for instance, nonrenormalizable operators are involved. Another possibility is that the field $X$ parametrizes a flat direction \cite{raby}. This is the case if in the superpotential the flat direction parametrized by the $X$ superfield is coupled to  some other superfield whose VEV is vanishing, {\it e.g.} $W=X\bar{\Phi}\Phi$, where $\bar{\Phi}+\Phi$ are a pair of vector-like superfields charged under some gauge group $G$. 
The $F$-component of the potential  $V(X)$ is vanishing and the flat direction   
  is lifted up by soft SUSY breaking terms   and  by loop-corrections. It is expected that $\langle X\rangle$ may assumed any value between the weak scale and 
the Planck scale \cite{raby}. Indeed,  at one-loop the potential of the $X$-field can be written as (for large values of the field)
\begin{equation}
V(X)= \widetilde{m}_X^2(Q) X^2+c F_X^2\ln\left(\frac{X}{Q}\right),
\end{equation}
where $\widetilde{m}_X^2$ is the soft SUSY breaking mass term evaluated at the scale $Q\simeq X$ and $c$ is a constant which depends on the degrees of freedom which couple to the superfield $X$. The soft SUSY breaking mass term
$\widetilde{m}_X^2$ may be  originated by supergravity corrections. Another possibility is that $\widetilde{m}_X^2$ receives contribution from one-loop Yukawa interactions. To illustrate this idea,  we can consider the following toy model

\begin{equation} W = \lambda_1  A \bar{\Psi}\Psi + B\left( \bar{\Psi}\Psi + \lambda_2
\Phi^+ \Phi^- + \lambda_3 B^2\right) 
 \end{equation}
where $A$ and $B$ are  singlets, $\Phi^{\pm}$ have 
 charge $\pm 1$ under a messenger $U(1)$  and $\bar{\Psi}$ and $\Psi$ are cherged  under 
some gauge group $G$. We assume that some  SUSY breaking occurs in a hidden sector 
dynamically and is transmitted  directly to the scalar states $\phi^{\pm}$ via 
the messenger $U(1)$ resulting in a negative mass squared $m^2$ for these two 
states. Minimizing the potential, one can show that there is a flat direction represented by $X\equiv \lambda_1 A+B$ whose VEV is undertermined at the tree-level and that supersymmetry is broken with $F_X = {m^2 \over \lambda_2} {1 \over (2 - \lambda_2/3 \lambda_3)}$. $\widetilde{m}_X^2$ gets a one-loop contribution proportional to $\lambda_2^2 m^2$ through  the Yukawa interaction $W=\lambda_2 B\Phi^+ \Phi^-$.

 Solving the renormalization group equations for   $\widetilde{m}_X^2(Q)$ one typically finds a solution of the form
\begin{equation}
\widetilde{m}_X^2(Q)\simeq \widetilde{m}_X^2(Q_0)\left[1+b\ln\left(\frac{Q}{Q_0}\right)\right],
\end{equation}
where $b$ is a coupling constant depending, again, on the degrees of freedom
which couple to the $X$. $\widetilde{m}_X^2(Q)$ goes through zero at $\overline{Q}\simeq Q_0\: {\rm e}^{-\frac{1}{b}}$. Since this mass term gives the dominant contribution to the effective potential for $X$, it is clear that $\langle X\rangle\sim \overline{Q}$ and, identifying $Q_0$ with $\Mpl$, it is reasonable to expect that $\langle X\rangle$ can take any value between the weak scale and $\Mpl$. As a result, $\sqrt{F_X}=\sqrt{\Lambda\langle X\rangle}$ can take any value between $10^4$ and $10^{12}$ GeV.

\section{The $\mu$-problem and one possible solution}
Usually the $\mu$-problem refers to the  the difficulty in generating the correct mass scale for the Higgs bilinear term in the superpotential
\begin{equation}
W_\mu =\mu H_U H_D.
\end{equation}
For phenomenological reasons, $\mu$ has to be of the order of the weak scale. In hidden $N=1$ supergravity it is possibile to generate $W_\mu$ if the K\"{a}hler potential has a non minimal form and if it is forbidden in the limit of exact supersymmetry \cite{gm}. The $\mu$-problem in the familiar GMSB theories appears at a first sight  more severe \cite{mu}. Indeed, it seems unnatural to have a $\mu$-term in the low energy theory since supersymmetry is broken dynamically and it would seem odd that the weak scale and the scale of $\mu$ coincide. Moreover, solutions existing in the framework of $N=1$ supergravity cannot be applied directly here since the SUSY breaking $F$-components are not usually  very large in GMSB models. In spite of these difficulties,  some solutions to the $\mu$-problem have already been proposed. In particular,  Leurer {\it et al.} have suggested a solution   which might be applied both in the context of  the $N=1$ supergravity scenario and in GMSB models \cite{le}. 

In addition to the usual  fields of the minimal supersymmetric standard model, there is another singlet, $S$. As a consequence of discrete symmetries, the coupling $S H_U H_D$ is forbidden in the superpotential. There are, however, various higher dimension couplings which can drive $\langle S\rangle$. In particular, consider terms in the effective Lagrangian of the form
\begin{eqnarray}
\label{s}
&&\int d^2\theta\left(
\frac{1}{\Mpl^p} X S^{2+p}-\frac{1}{\Mpl^m} S^{m+3}+\frac{\alpha}{\Mpl^n}S^{n+1}H_U H_D\right)\nonumber\\
&+& \frac{1}{\Mpl^2}\int d^4\theta\:X^{\dagger}X S^{\dagger}S.
\end{eqnarray}
This structure can be enforced
by discrete symmetries. 

In the $N=1$ supergravity scenario the superfield $X$ may be interpreted to be part of the hidden sector and therefore $\sqrt{F_X}\sim 10^{11}$ GeV.   In the context of   GMSB models,  it may be the same singlet superfield responsible for the splitting in the  spectrum of the messangers $(q,\ell)$.

The first and the fourth terms in Eq. (\ref{s}) can contribute to the effective negative curvature terms to the $S$ potential. If $\langle X\rangle\ll \Mpl$ and, for example,  $p=m=2$ and  $n=1$,  the $\mu$-term turns out to be 
\begin{equation}
\mu\simeq \alpha \sqrt{F_X},
\end{equation}
Since the operator  $\frac{\alpha}{\Mpl^n}S^{n+1}H_U H_D$ in the superpotential  is expected to arise in the effective theory after having integrated out some heavy fields, the coefficient $\alpha$ is expected to be very small.  It will be  equal to some  powers of coupling constants times,  eventually,  some ratio of mass scales. For the mechanism to work, it is required that $\alpha\sim 10^{-7}$ or so.

Besides generating a $\mu$-term, this mechanism can also give rise to a nearly
vanishing  $B_\mu$-term, {\it i.e.}
 the soft supersymmetry breaking
term $B_\mu H_uH_d$ in the Higgs potential.
 It is noticeable that  boundary conditions equal to zero for
 bilinear (and trilinear) soft parameters at
the messanger scale makes
the GMSB models free from the supersymmetric $CP$ problem and highly predictive \cite{bor}. 

What is crucial for us is that the field $S$, although very weakly coupled to ordinary matter, may play a significant role in cosmology. We will devote the rest of the paper to explore the cosmological implications of such a field and to show that a succesful inflationary scenario may be constructed out of the potential for the field $S$. The fact that the inflaton field may be identified  with the same scalar responsible for the generation of the $\mu$-term satisfies the minimal principle and  allows us to connect a theory which could be tested at accelerators
with measurements of the  temperature anisotropy in the CMB. 

\section{The inflationary stage}

Let us suppose that the phase transition during which the  $X$-field acquires a vacuum expectation value   for both its scalar and auxiliary components  takes place at temperatures of order of $\sqrt{F_X}$ or higher and that, both in the case in which   SUSY is mediated by gravitational interactions and by gauge forces supersymmetry breaking is parametrized by an $F$-term of the order of $F_X$. 
  
Let us  restrict ourselves to the case  $p=m=2$, and  $n=1$. The potential along the real component of the  field $S$ reads
\begin{eqnarray}
\label{global}
V(S) &\sim & F_X^2-\frac{F_X^2}{\Mpl^2}S^2-\frac{F_X}{\Mpl^2}S^4\nonumber\\
&+&  \frac{X^2}{\Mpl^4}S^6 - \frac{X}{\Mpl^4} S^7+\frac{1}{\Mpl^4}S^8,
\end{eqnarray} 
Under the condition $\langle X\rangle \ll \Mpl$,  the true vacuum is at
\begin{equation}
 \langle S\rangle^4\sim  F_X \Mpl^2,
\end{equation}
 such that the $\mu$-term is proportional to  $\langle S\rangle^2/\Mpl\sim \sqrt{ F_X}$. Notice that 
we have added the constant $\sim F_X^2$ in such  a way that the the cosmological constant in the true vacuum is zero,   $V(\langle S\rangle)=0$.

Around $S=0$ we may considerably  simplify the potential as
\begin{equation}
V(S)\simeq V_0-\frac{m^2}{2}S^2-\frac{\lambda}{4}S^4,
\end{equation}
where
\begin{equation} 
V_0 \sim  F_X^2,~~~~~  m^2\sim 2 \frac{F_X^2}{\Mpl^2} ~~{\rm  and} ~~ 
\lambda\sim 4 \frac{F_X}{\Mpl^2}.
\end{equation}
If the $S$-field starts sufficiently close to the origin the system may inflate. 

It is important to notice that 
the  potential is characterized by a drastic steepening of the quadratic term. 
This means that, during inflation and  soon after cosmological scales leave the horizon, the quartic term starts dominating. 
The quadratic and the quartic terms  become comparable for $S_*\sim\sqrt{F_X}$. Since this value   is much smaller than $\langle S\rangle$ and  all the dynamics giving rise to density perturbations
takes place  in the vicinity of the origin, the presence of the quartic term 
cannot be neglected as usually done  in the   determination of the CMB anisotropy and the spectral index of the power spectrum \cite{lr}.

Before launching ourselves into the peculiar features of the model 
it is  useful to deal with some generalities. During inflation the
potential $V(S)$ is supposed to 
satisfy the flatness conditions $\epsilon\ll 1$ and $|\eta|\ll1$, 
where
\begin{eqnarray}
\epsilon&\equiv &\frac12\Mpl^2(V'/V)^2,\\
\eta&\equiv &\Mpl^2 V''/V.
\end{eqnarray}
Given these conditions, the evolution of the $S$-field  
\be
\ddot S+3H\dot S=-V'
\label{phiddot}
\ee
typically settles down to the slow roll evolution
\be
3H\dot S=-V', 
\label{slowroll}
\ee
where $H=\sqrt{\frac{V_0}{3\Mpl^2}}$ represents   the Hubble parameter during inflation. 

Slow roll conditions are motivated by the observed fact that the spectrum has mild scale
dependence. Moreover, 
slow roll and the flatness condition $\epsilon\ll 1$
 ensure that the energy density 
$\rho_S=V(S)+\frac{1}{2}\dot{S}^2$ is close to $V$ and is slowly varying.
As a result $H$ is slowly varying,
which implies that the scale factor $a$ of the universe grows exponentially,  
$a\propto {\rm e}^{Ht}$ at least over a Hubble time or so.The
flatness condition $|\eta|\ll 1$ then ensures that 
$\dot S$ and $\epsilon$ are slowly
varying.

A crucial role is played by the number of Hubble times
$N(S)$ of inflation, still remaining when $S$ has a given value.
By definition $dN=-H\,dt$, and 
the 
slow roll condition together with the flatness condition
$\epsilon\ll 1$ lead
\be
N=\left| \int^S_{S_{\rm end}} \Mpl^{-2}\frac V{V'} d S
\right|
\label{nint}
\ee
If we assume that the quadratic term dominates while cosmological scales are leaving the horizon, the slow roll parameter $\eta$ is given by $\eta=\frac{m^2 \Mpl^2}{V_0}$. 
Since the slow roll paradigm is well motivated, while cosmological 
scales are leaving the horizon, by the observed fact that 
the power spectrum of density perturbation does not vary much on such scales, a fundamental question is whether the slow roll conditions are satisfied in the model we are discussing. 

\subsection{The $\eta$-problem}
When dealing with inflation model building in the context of supersymmetric theories one has always to face a serius problem. The  generalization of supersymmetry from a global to 
a local symmetry automatically incorporates gravity and, therefore, inflation model building must be considered in the framework of supergravity theories. In other words, the potential (\ref{global}) should  be extended to incorporate supergravity effects, see Eq. (\ref{sugra}). This obviously  holds  in the case in which  SUSY breaking is transmitted by gauge interactions, but it is also true in the context of GMSB models. 

In small-field models of inflation (values of fields smaller than the reduced Planck scale $\Mpl$), where the theory is under control,
it is reasonable to work in the context of 
supergravity. 
The supergravity potential is rather involved, but it  can still  be written as 
a $D$-term plus an $F$-term, and it is usually supposed that
the $D$-term vanishes during inflation. 
Now, for models where the $D$-term vanishes,
the slow roll  parameter $\eta=\Mpl^2 V''/V$ generically receives
various contributions of order $\pm 1$. This is the so-called $\eta$-problem of supergravity theories.  This  
crucial point was first emphasized in Ref.~\cite{CLLSW}, though it
is essentially 
a special case of the more general result,
noted much earlier
\cite{dinefisch,coughlan}, that 
there are contributions of order $\pm H^2$ to the mass-squared of every
scalar field. Indeed, in a small-field
model the troublesome contributions to $\eta$ may
be regarded as contributions
to the coefficient $m^2$ in the expansion  of the inflaton
potential. Therefore, it is very difficult to naturally implement a slow roll inflation in the context of supergravity. The problem basically  arises since inflation, by definition, breaks global supersymmetry because of a nonvanishing cosmological constant (the false vacuum energy density of the inflaton). In supergravity theories, supersymmetry breaking is transmitted by gravity interactions and the squared mass of the inflaton becomes naturally of order of $V/\Mpl^2\sim  H^2$. The perturbative renormalization of the K\"ahler potential is therefore crucial   for the inflationary dynamics due to a non-zero energy
density which breaks supersymmetry spontaneously during inflation and usually it is not temable.  

 Even though it is beyond the scope of this paper to propose a solution to the $\eta$-problem, we would like to point out that how severe the problem is depends on the magnitude of $\eta$ \cite{lr}.
If $\eta$ is 
not too small then its smallness could be due to accidental 
cancellations. On the other hand,  having $\eta$ not too small requires 
that the spectral index  $n=1-6\epsilon+2\eta$ is significantly displaced from 1. It is noticeable that accidentale cancellations giving rise to small values of $\eta$ are not inconceivable in the model discussed in this paper. Indeed, supergravity contributions to $\eta$ coming from the K\"{a}hler potential  
may  be cancelled by the  term of the form $S^{\dagger}S X^{\dagger}X$ in the lagrangian. As we shall see in the following, a peculiar prediction of the model is that the spectral index may be significantly displaced from 1. This tells us that accidental cancellations are not so unlikely in the present context .

\subsection{The predictions of the power spectrum and the spectral index}

The quantum fluctuation of the inflaton field
gives rise to an adiabatic density
perturbation, whose spectrum is
\be
\delta_H^2(k) = \frac1{75\pi^2\Mpl^6}\frac{V^3}{V'^2}
=
\frac1{150\pi^2\Mpl^4}\frac{V}{\epsilon}.
\ee
In this expression, the
potential and its derivative are evaluated at the epoch of 
horizon exit for the scale $k$, which is defined by $k=aH$. 
The COBE measurement  gives an accurate determination of 
$\delta_H$ on the corresponding scales because the evolution is
purely gravitational (dominated by the Sachs-Wolfe effect).
On the
scale $k\simeq 5H_0$ one finds \cite{delh,bunnwhite}
\be
\delta_H= 1.94\times 10^{-5}, 
\label{cobe}
\ee
with a $2\sigma$  uncertainty of $15\%$.
This assumes that gravitational waves give a negligible contribution.

Comparison
of the prediction
 with the value deduced from the COBE observation of the CMB 
anisotropy gives 
\be
\Mpl^{-3} V^{3/2}/V' = 5.3\times 10^{-4}.
\label{cobecons}
\ee
This relation provides a useful constraint
on the parameters of the potential. In our case, $V' =- m^2 S -\lambda S^3 + \cdots$ and the 
 two terms are equal at $S\simeq S_*\equiv m/\sqrt\lambda\sim \sqrt{F_X}$. If we 
suppose that the first term dominates while cosmological
scales are leaving the horizon, but that the second term dominates 
before the end of inflation, it is easy to show that   
\be
\frac{S}{S_*}\simeq \exp\left(\frac12 - x\right),
\ee
where $x\equiv \frac12N(1-n)$. Consistency with the assumptions made imposes that  $x >\frac12$.

With fixed $n$ and $N$, the COBE
normalisation determines $\lambda$ to be 
\be
\lambda =  2\times 10^{-13} \left(\frac{50}{N}\right)^3 (2x)^3 {\rm e}^{(1-2x)}.
\label{cobe}
\ee
This COBE normalized  value of $\lambda$ is  smaller than the corresponding value
for a pure $S^4$ model with a potential of the form $V=V_0-\frac{\lambda}{4}S^4$  \cite{lr}. 
With the minimum value $x=\frac{1}{2}$, one reproduces the pure $S^4$ result \cite{lr}, otherwise $\lambda$ is smaller.  The amplitude of the gravitational waves produced by quantum fluctuations is far too small to be detected since the variation of the field during inflation is much smaller than $\Mpl$ \cite{grav}.

\subsection{The scale of the vacuum energy}

The COBE normalized value of $\lambda$ 
 allows us to fix the scale $\sqrt{F_X}$:  
\begin{equation}
\sqrt{F_X}\simeq 5.4\times 10^{11} \left(\frac{50}{N}\right)^{3/2} (2x)^{3/2} {\rm e}^{(\frac{1}{2}-x)}\:\: {\rm GeV}.
\end{equation}
 The exact value  of $N$ at which cosmological scales leave the horizon   can  be 
determined if the history of the universe after inflation 
is known.
Consider  the epoch when the 
scale
$k^{-1}=H_0 ^{-1}\sim 3000 h^{-1}$ Mpc leaves the horizon, which can be 
taken to mark the 
beginning of cosmological inflation. Using a subscript 1 to denote this epoch, $N_1= \ln(a_{\rm end}/a_1)$, where the subscript ``end" denote the end of inflation,  is given by  \cite{LL2}
\be 
\label{n1}
N_1=62-\ln(10^{16}\GeV/V_{\rm end}^{1/4}) 
-\frac13\ln(V_{\rm end}^{1/4}/\rho_{\rm reh}^{1/4}).
\ee
This formula assumes  that the end of inflation gives way promptly to
matter domination, which is followed by a radiation dominated era
lasting until the present matter dominated era begins. 
$\rho_{\rm reh}^{1/4}$ is the reheating temperature when radiation domination 
begins. In the following we will see that the reheating temperature $\rho_{\rm reh}^{1/4}$ is of order of 1 GeV or so and, since the dependence of $N_1$ is only logarithmically dependent on $V_{\rm end}^{1/4}$, a reasonable value for $N_1$ is given by $N_1\simeq 40$. 

As far as the spectral index is concerned, the   four year COBE measurement gives $n=1.2\pm 0.3$ at the 1$\sigma$ level \cite{four}. 

The precise determination of the spectral index $n$  involves measurements of $\delta_H$ also on small scales. The 
  main uncertainties
are the value of the Hubble constant 
$H_0=100h\km {\rm s}^{-1}{\rm Mpc}^{-1}$,
the value of the 
baryon density $\Omega_B$, and the 
nature of the dark matter. For instance, the   case of  pure cold dark matter is  viable at 
present \cite{constraint2} with $h\simeq 0.5$, $\Omega_B\simeq 0.12$ and the spectral index is constrained to be in the range
\be
0.7 \lsim n \lsim 0.9
\ee
If there is  an admixture
of hot dark matter in the form of a single neutrino species
\cite{mdm},  and taking
$\Omega_B<0.15$ and $h>0.4$, the lower bound does not change significantly, but $n$ is bounded from above to be smaller than about 1.3.  

As we have mentioned previously, it is more natural that in the model discussed in this paper the spectral index is significantly smaller than 1,  leading to the so-called  red spectrum. If so, accidental cancellations in the expression for $\eta$ between the supergravity contributions  and term   of the form $S^{\dagger}S X^{\dagger}X$ in the lagrangian  become more natural. On the other hand, the COBE normalized value of $\sqrt{F_X}$ is very sensitive to  the value of the spectral index. 

For $n=0.7$ we find  $\sqrt{F_X}\simeq 1.2\times 10^{11}$ GeV which is exactly the scale required in the framework of $N=1$ hidden supergravity. This is   quite an intriguing coincidence. It is also remarkable that the energy scale is much smaller than the one required in many alternative inflationary models \cite{lr}, usually of the order of the grand unified scale. This is a result of the the  drastic steepening of the quadratic term in the potential $V(S)$.

If we allow a lower value of the spectral index,  $n=0.6$, which is still consistent at 2$\sigma$ level  with the four year COBE measurement,  we find $\sqrt{F_X}\simeq 1.1 \times 10^{10}$ GeV. This value is slightly too small to be consistent with the supergravity scenario since  superpartners would have masses of the order of 50 GeV.  At a first sight this value might seem too large even in the framework of GMSB models. However, as we explained in the section II, the spectrum of the superparticles only fixes    the  ratio $\Lambda= F_X/X$  to be relatively small and around $10^3$ TeV, while $\sqrt{F_X}$ may be much larger than $\Lambda$\footnote{One should also note that
a SUSY breaking scale as high as $10^{11}$ GeV does not imply necessarily that  
gravitational interactions are the only mediators of SUSY breaking. One can always envisage a mixed scenario in which the soft breaking masses of the sfermions and gauginos receive contributions from both gravitational and gauge interactions. In this paper, however, we take  the attitude that the two sources do not overlap.}.    

A spectral index $n$ larger than about 0.8 gives a COBE normalized scale of SUSY breaking larger than about $3 \times 10^{11}$ GeV, with a corresponding gravitino mass $m_{3/2}=\frac{F_X}{\sqrt{3}\Mpl}\gsim 10$ TeV. This latter value seems  slightly too high to be consistent with naturalness arguments suggesting  that sfermion masses should be lighter than about 1 TeV or so. 

All these considerations lead us to conclude that the model studied in this paper is quite predictive.  The density perturbations generated during the inflationary stage driven by the field $S$   should be  characterized by a spectral index $n$ in the range $(0.6-0.8)$.   Moreover,  larger values of $|n-1|$ are associated to relatively smaller values of the SUSY breaking scale. 

 The next measurements of the temperature anisotropies in the cosmic microwave background   will  confirm or disprove  these expectations. If the future observations will indicate a  value of the spectral index significantly displaced from 1, this might be interpreted as a signal that Nature has chosen the same scale at which breaking supersymmetry and driving inflation. 
It is intriguing that the next measurements of the two-point correlation function of the temperature anisotropy may help us to understand how large  is the scale of supersymmetry breaking,  what  is the mechanism which mediates it,   and to  get a depper insight into the $\mu$-problem.

\subsection{The problem of initial conditions}

Before studying in details the post-inflationary era, let us briefly address the issue of the initial condition for the field $S$.  We have assumed that underlying the
model are discrete symmetries under which $S$ transforms non-trivially.  As a result,
$S=0$ is a special point, and it is natural that $S$ may sit at this point initially.
This despite the fact that it is very weakly coupled to ordinary matter, and might not
be in thermal equilibrium. 
Many models of slow rollover inflation require a fine-tuning in the initial value for the field to be successful and the smaller is the scale of inflation the more severe is the fine-tuning \cite{tu}. From Eq. (\ref{nint}) we may infer that,  in order to achieve the 40 or so {\it e}-foldings of inflation required, the initial value of the scalar field must be less than about $2\times 10^{7} \left(\sqrt{F_X}/10^{10}\:{\rm GeV}\right)$ GeV.
As a result, only regions  where the initial value of the field is small enough will undergo inflation. These regions have grown exponentially in size and they should occupy most of the physical volume of the Universe.

 We notice that the small value of the field  is not spoiled by quantum fluctuations which are of the order of 
\begin{equation}
\frac{H}{2\pi}\sim 5 \left(\frac{\sqrt{F_X}}{10^{10}\:{\rm GeV}}\right)^2\:\:{\rm  GeV}.
\end{equation}

 Thermal fluctuations might spoil such a localization since $
\langle S^2\rangle_T^{1/2}$ would be naturally of the order of $ T\sim \sqrt{F_X}$. However, the inflaton field is so weakly coupled, being  its
couplings  all suppressed by powers of $\Mpl$,  that it is easy to check that  thermal contact with the rest of the Universe has never been established \cite{tu,ross}.

A possible dynamical tuning of the initial condition for the field $S$ may be implied by a short period of ``pre-inflation" \cite{dyn}\footnote{We assume here that density perturbations generated at this epoch are negligible.}. Indeed, let us imagine that the universe underwent a short period of inflation with Hubble parameter $H_{{\rm p-i}}$ before $X$ and $F_X$ acquire a vacuum expectation value. In such a case the field $S$ gets an effective mass of the order of  $H_{{\rm p-i}}$ from supergravity corrections and  it oscillates around $S=0$ with its amplitude decreasing as $a^{-3/2}$. At the end of the pre-inflationary stage the $S$ takes the value $S_{{\rm p-i}}\simeq S_i \:{\rm e}^{-\frac{3}{2}N_{{\rm p-i}}}$, where $S_i$ is the value of $S$ at the beginning of pre-inflation and $N_{{\rm p-i}}$ is the number of $e$-folds relative to pre-inflation. If we take $S_i\sim \Mpl$, $\sqrt{F_X}\sim 10^{10}$ GeV  and require that 
\begin{equation}
S_{{\rm p-i}}\lsim S_* \exp\left(\frac12 - x\right),
\end{equation}
we find 
\begin{equation} 
N_{{\rm p-i}}\gsim 20.
\end{equation}
 Quantum fluctuations generated during the pre-inflationary era are of the order of $\frac{H_{{\rm p-i}}}{2\pi}\:{\rm e}^{-\frac{3}{2}N_{{\rm p-i}}}$ and do not kick the condensate to values larger  than $S_*{\rm exp}\left(\frac{1}{2}-x\right)$. Therefore, if the ordinary period of inflation was preceded 
by a short period of pre-inflation \cite{dyn}, it is reasonable to expect that the field $S$ is  sitting with great accuracy so close to the origin  that  ordinary inflation may successfully take  place. 

\section{The post-inflationary stage}

Let us now consider  the dynamics of the inflaton field after the end of inflation. After its slow roll, the field $S$ begins to oscillate about the minimum of its potential and the vacuum energy that drives inflation is coverted into coherent scalar field oscillations corresponding to a condensate of nonrelativistic $S$-particles \cite{kt}. During this epoch of  coherent $S$-oscillations the universe is matter dominated and the energy trapped in the $S$-field decreases as $a^{-3}$. 
The conversion of the vacuum energy to thermal radiation, usually dubbed reheating, takes place when the $S$-particles decay into light fields, which, through their decays and interactions produce a thermal bath. The reheating temperature is determined by the decay time of the scalar field oscillations
which is set by the inverse of the decay width $\Gamma_S$ of the field $S$. If $\Gamma_S\gsim H$, the $S$-oscillations decay rapidly and the vacuum energy is entirely converted into radiation corresponding to a very high value of the reheating temperature, $T_{{\rm RH}}\simeq \sqrt{F_X}$. However, 
 if $\Gamma_S\lsim H$, which is the rule in slow roll inflation, the coherent oscillation is relatively long and the reheating temperature turns out to be of order of 
\begin{equation}
T_{{\rm RH}}\simeq 0.5\:\sqrt{\Gamma_S\Mpl}\ll\sqrt{F_X},  
\end{equation}
corresponding to a partial conversion of vacuum energy into radiation.  

At the minimum of its potential, the scalar field has a mass
 
\begin{equation}
m_S=\sqrt{V^{\prime\prime}(S)}\simeq 3 \frac{F_X^{3/4}}{\Mpl^{1/2}}\simeq
10^3\left(\frac{\sqrt{F_X}}{10^{10}\:{\rm GeV}}\right)^{3/2}\:{\rm TeV}.
\end{equation}
The scalar oscillations may decay into light Higgsinos $S\rightarrow \widetilde{H}_U\widetilde{H}_D$ with a rate
$\Gamma_S=\frac{g^2 m_S}{4\pi}$ where 
\begin{equation}
g\sim\alpha \frac{\langle S\rangle}{\Mpl}\sim\frac{\mu}{\sqrt{F_X}}\frac{\langle S\rangle}{\Mpl}
\end{equation}
and we have properly  taken into account the fact that  $\mu\ll \sqrt{F_X}$. The  resulting reheating temperature is then 
\begin{equation}
T_{{\rm RH}}\sim 10^{-2} \mu F_X^{1/8}\Mpl^{-1/4}\simeq 5\times 10^2\left(\frac{\mu}{
10^3\:{\rm GeV}}\right)\left(\frac{\sqrt{F_X}}{
10^{10}\:{\rm GeV}}\right)^{1/4} \: {\rm MeV}.
\end{equation} 
$T_{{\rm RH}}$ is   large enough to preserve the classical cosmology beginning with the era of nucleosynthesis. It seems difficult, however, to push the reheating temperature above the electroweak scale, $T_{{\rm EW}}\sim 100$ GeV. Thus, it appears that supersymmetric electroweak baryogenesis \cite{bau} is not a viable option
for the generation of the baryon asymmetry. On the other hand,
the decays of the inflaton themselves might be responsible for the baryon asymmetry \cite{b}.  

\subsection{The generation of the baryon asymmetry}

Let us imagine that the  couplings by which the inflaton decays may contain $CP$-violation and baryon number violation. In order to produce a baryon asymmetry, we must have baryon number violating operators in the Lagrangian, such as 
\begin{equation}
\delta W\sim \lambda\left(\frac{S}{\Mpl}\right)\bar{U}\bar{D}\bar{D},
\end{equation} 
where generation indeces are suppressed
The presence of such operator is compatible with the stability of the proton and the experimental absence of neutron-antineutron oscillations \cite{n}. The baryon number violating decay rate is given by \cite{thomas}
\begin{equation}
\Gamma_B\simeq\frac{6\bar{\lambda}^2}{(8\pi)^3}\:\frac{m_S^3}{\Mpl^2},
\end{equation}
where  $\bar{\lambda}^2\equiv \sum \left|\lambda\right|^2$ is the sum over the generations of the final state. We can estimate the baryon asymmetry produced by the inflaton decay in the following way. 

We assume that the amount of baryon number produced per decay is $\varepsilon$. $\varepsilon$ is the product of $CP$-violating phases $\delta_{CP}$ times some loop factors times the ratio of the baryon number violating decay rate over the total decay rate 
\begin{equation}
\frac{\Gamma_{B}}{\Gamma_{{\rm tot}}}\sim 10^{-2}\:\bar{\lambda}^2\: \left(\frac{\mu}{\sqrt{F_X}}\right)^{-2}\left(\frac{m_S}{\langle S\rangle}\right)^2.
\end{equation} 
The number of massless particles produced per decay is $\sim \frac{m_S}{T_{{\rm RH}}}$. Plugging in the expected values  of the inflaton mass and the reheating temperature for $\sqrt{F_X}\sim 10^{10}$ GeV, we find a baryon to entropy ratio 
\begin{equation}
B\simeq 10^{-1}\times ({\rm loop}\:\:{\rm factors})\times \delta_{CP}\:\bar{\lambda}^2,
\end{equation}
which can account for the observed baryon asymmetry $B\sim 10^{-10}$ for $\delta_{CP}\sim \bar{\lambda}\sim 10^{-2}$. 

\subsection{The fate of gravitinos and moduli}

Let us finally discuss the cosmology of gravitinos and moduli when a period of late inflation takes place. 

As we have  mentioned in the introduction, in GMSB models gravitinos are usually expected to be lighter than what predicted in the framework of $N=1$ supergravity theories, since the mass of the gravitino is fixed by the scale of the $F$-term which breaks supersymmetry, $m_{3/2}\sim \frac{F}{\Mpl}$. However, if we insist in identifying  the scale $F_X$ with the one suggested by COBE normalization, gravitinos are not dramatically  light,  $m_{3/2}\simeq 50$ GeV. These relatively  heavy gravitinos do not pose any cosmological problem. 

In GMSB models  the gravitino may be  the lightest supersymmetric particle. Since $R$-parity is supposed to be broken only  in the baryon number operator $\delta W\sim \lambda\left(\frac{S}{\Mpl}\right)\bar{U}\bar{D}\bar{D}$, it is easy to show that the gravitino lifetime is longer than the present age of the universe so that the gravitino can be considered  cosmologically stable. 
If a stable gravitino is thermalized in the early universe and not diluted by any mechanism, its mass density may exceed the closure limit $\Omega_{3/2}\lsim 1$. Since the number density of gravitinos is fixed once they  are thermalized, the above argument sets a stringent  upper bound on the gravitino mass,  $m_{3/2}\lsim 2$ keV when  no source of  dilution is present  \cite{pargel}. 
However, gravitinos are efficiently diluted during the inflationary stage driven by the field $S$ and they are not produced in the subsequent stage of reheating. Indeed,   gravitinos  may be regenerated during reheating  either by the  decays of sparticles (or particles in the messanger sector)    or by scatterings processes \cite{murayama1,murayama2}. The first mechanism requires the reheating temperature to be at least of order of  the typical sparticle mass,  $\widetilde{m}\sim$ 100 GeV. Since the reheating temperature is at most of order of 1 GeV, the production of heavy states is drastically  suppressed and gravitinos are not produced by decays of sparticles. Scattering processes are much more dangerous. Gravitinos are produced more at higher temperatures, which provides an upper bound on the reheating temperature from the bound  $\Omega_{3/2}\lsim 1$:  for $m_{3/2}\simeq 50 $ GeV, one gets $T_{{\rm RH}}\lsim 10^9$ GeV \cite{murayama1}. This bound is  satisfied   in the scenario depicted so far where  
inflation takes place at late times. We may safely conclude that stable  gravitinos were not populating the  universe at the beginning of the radiation era:   the  stable gravitino problem is solved by the late stage of inflation and by the fact that the reheating temperature is so low. 

This situation is   much different from the one depicted   in ref. \cite{murayama2}. There it was assumed that  the scale of supersymmetry breaking  is  much smaller than  $10^{10}$ GeV and, therefore, the gravitino is very light. It was also supposed that a primordial stage of inflation is terminated by reheating the universe up to a temperature of order of $10^8$ GeV or higher. Under these circumnstances , it was concluded that stable gravitinos are  inevitably regenerated in great abundance    during the reheat stage and that a significant amount of entropy release must take place after inflation to diluite them.  This large amount of entropy release is not necessary in the scenario analyzed here since stable gravitinos are not populating the universe after reheat.

If the gravitino is not the lightest supersymmetric particle, it may decay with a typical lifetime $\tau_{3/2}\sim 10^{2}\frac{\Mpl^2}{m_{3/2}^3}$. This certaily occurs when  supersymmetry breaking is mediated by gravitational interactions and the gravitino may be as heavy as 1 TeV, but it is also a possibility in the framework of GMSB models.  

Decays occur after the big-bang nucleosynthesis and produce an unacceptable amount of entropy, which conflicts with the prediction of  big-bang nucleosynthesis. In order to keep the success of big-bang nucleosynthesis, the gravitino mass should be larger than about 10 TeV. However, if the universe went through a period of late inflation, any initial abundance of gravitino is diluited by the exponential expansion of the universe, but gravitinos are regenerated during the reheatings stage.   The most stringent upper bound on $T_{{\rm RH}}$ in the case of unstable gravitinos comes from the photo-dissociation of light nuclei. Indeed, if gravitinos decay radiatively, the emitted high energy photons induce cascade processes and affect the results of big-bang nucleosynthesis. Other possible constraints are from the mass density of the lightest supersymmetric particle and the enhancement of cosmic expansion due to the gravitino. A detailed analysis has been performed in \cite{nucleo} where it was concluded that grav
itinos in the mass range $(10^2-10^3)$ GeV are harmeless for reheating temperature smaller than about  $(10^6-10^8)$ GeV. Again, this is satisfied in the scenario studied in this paper. 

If we assume that the underlying theory is a string theory, we have also to face the so-called 
 cosmological moduli problem \cite{moduli}.  In string models massless fields exist in all known string ground states and parametrize the continuous ground state degeneracies characterisitic of supersymmetric theories. 
These fields ${\cal M}$ are massless to all orders in perturbation theory and get their mass of order the weak scale from the same  mechanism which breaks supersymmetry.  Being coupled to the ordinary matter only by gravitational strength, a dangerously long lifetime results. Indeed,  if one of these fields at early epochs is sitting far from the minimum of its potential with an amplitude of order of the   Planck scale,  the  coherent oscillations about the minimum  will    eventually dominate the energy density of the universe.    These fields will then behave like nonrelativistic matter and decay at very late times, dominating the energy of the universe until it is too late for nucleosynthesis to occur.  A related and possibly more serious problem is that, during the decays, an enormously amount of entropy is released erasing out any pre-existing baryon asymmetry. 

This problem is
somewhat ameliorated in our scenario. 
 The  period of inflation  driven by
the field $S$ may take place
at a sufficiently   late stage of the universe,   $H\lsim m_{{\cal M}}$, that
the number density of string moduli is reduced by a factor ${\rm exp}(-3N)$ and  by the subsequent entropy production at the reheat stage \cite{randall}. It is an attractive
feature of the present scenario that this is possible.
However, this is not generally enough, since the minimum
of the moduli potential, generically, will be shifted by an amount of order $\Mpl$ during
inflation, as a result
of couplings of the moduli
\begin{equation}
\int d^4 \theta\left[ X^{\dagger} X f\left(\frac{{\cal M}}{\Mpl}\right) + S^{\dagger} S
g\left(\frac{{\cal M}}{\Mpl}\right)\right].
\end{equation}
So it is probably necessary to invoke symmetry reasons
so that the minima  during inflation and at the present day coincide  to a high degree of accuracy.
In the  context of GMSB models, 
it might be tempting   to abandon the assumption that $\sqrt{F_X}$ is as high as $10^{10}$ GeV and to concern ourselves with smaller values of $\sqrt{F_X}$,  of the order of $(10^5-10^6)$ GeV. In such a case, it is clear that the inflationary stage driven by the field $S$ cannot be responsible for the generation of the density perturbations, but it might be useful to diluite light string moduli. A number of $e$-folds larger than about 5 would be sufficient to diluite sting moduli by a factor $10^{-15}$.  It is also easy   to check  the number of $e$-folds cannot exceed $\sim 20$ or so in order to keep the primordial density fluctuations generated by a ``standard" inflation with $H\sim 10^{13}$ GeV \cite{randall}.  By looking at Eq. (\ref{n1}), we realize that  in the present model this is not a viable option. Indeed, the number of $e$-folds turns out to be simply too  high, $N_1\sim 30$.

\section{Conclusions and outlook}

One of the most important paradigms in the cosmology of the early universe is that the  latter suffered a period of accelerated expansion. This inflationary stage provides a possible solution to cosmological conundrums
such as the flatness, the horizon and the monopole problems. It is generaly believed that any successful inflationary scenario is intimately connected to some new physics at extremely high scales. Observational cosmology is now entering  a new and exciting period where it is becoming possible to test  inflationary models for the first time. Such  new measurements represent a unique occasion to get some insight into  new physics beyond the standard model.  

Inflation, as currently understood, requires the presence of fields with
very flat potentials.  The extreme flatness of the inflaton potential seems technically
natural only in supersymmetric models since the nonrenormalization theorems guarantee that the flateness of the potential is not spoiled by radiative corrections.  Therefore, it seems that the way Nature has chosen to accomplish inflation depends upon the 
way and the scale at which supersymmetry is broken.  Two different ways of transmitting supersymmetry breaking are currently very popular: in the  supergravity scenario it is gravity which acts as mediator and in GMSB
models standard model this role is played by gauge interactions.   

By taking inspiration from a possible solution to the $\mu$-problem, we have presented  a simple model for slow rollover inflation which is minimal in the sense that the inflaton may be  identified with the  field responsible for the generation of the $\mu$-term. Inflation takes place when the inflaton condensate is rolling down from the origin and the potential is characterized by a dramatic steepening of the quadratic term. 
This implies that  the  COBE normalized value of the vacuum energy  may be  naturally of the order of $10^{11}$ GeV which seems to favor the $N=1$ supergravity scenario.   Smaller values of the vacuum energy are not presently excluded and therefore the  inflationary scenario may be also implemented 
in the context of GMSB models. This option is pleasing since in this class of models supersymmetry  breaking masses are proportional to the $F$-term of a gauge singlet superfield and 
the same  $F$-term  may provide the vacuum energy density necessary to drive inflation. 

We have shown that the  reheating after the end of inflation is not very efficient. The particles popping out from the decay of the inflaton oscillations around the true minimum of the potential rapidily thermalize with a typical energy  $T_{{\rm RH}}\sim 1$ GeV. As a result, gravitinos cannot be  produced by thermal scatterings and are cosmologically harmless.  The number density of string moduli may be  reduced during late inflation, however a detailed inspection of the string moduli potential is needed before making any   solid prediction. In spite of the low reheating temperature, the baryon asymmetry may be generated by the decays of the inflaton field if the latter has nonrenormalizable and baryon number violating couplings.

 Finally, the spectrum of density perturbations is characterized by a spectral index which is significantly displaced from 1. This does not come as a surprise since large values of $|n-1|$ are generally associated to relatively small values of the vacuum energy. This prediction  has  the  advantage that
the next measurements of the temperature anisotropies in the CMB will be able to say the last word about the viability of the model and, hopefully, on the way supersymmetry is fed down to the observable sector.

\vskip 1cm
\centerline{\bf Acknowledgements}

A.R. would like to thank   M. Dine for an enjoyable  collaboration,  for many illuminating discussions and for a careful reading of the manuscript. D.H. Lyth and W. Kinney are  also acknowledged for several comments. 
A.R. supported by the DOE and NASA under Grant NAG5--2788.

\def\NPB#1#2#3{Nucl. Phys. {\bf B#1}, #3 (19#2)}
\def\PLB#1#2#3{Phys. Lett. {\bf B#1}, #3 (19#2) }
\def\PLBold#1#2#3{Phys. Lett. {\bf#1B} (19#2) #3}
\def\PRD#1#2#3{Phys. Rev. {\bf D#1}, #3 (19#2) }
\def\PRL#1#2#3{Phys. Rev. Lett. {\bf#1} (19#2) #3}
\def\PRT#1#2#3{Phys. Rep. {\bf#1} (19#2) #3}
\def\ARAA#1#2#3{Ann. Rev. Astron. Astrophys. {\bf#1} (19#2) #3}
\def\ARNP#1#2#3{Ann. Rev. Nucl. Part. Sci. {\bf#1} (19#2) #3}
\def\MPL#1#2#3{Mod. Phys. Lett. {\bf #1} (19#2) #3}
\def\ZPC#1#2#3{Zeit. f\"ur Physik {\bf C#1} (19#2) #3}
\def\APJ#1#2#3{Ap. J. {\bf #1} (19#2) #3}
\def\AP#1#2#3{{Ann. Phys. } {\bf #1} (19#2) #3}
\def\RMP#1#2#3{{Rev. Mod. Phys. } {\bf #1} (19#2) #3}
\def\CMP#1#2#3{{Comm. Math. Phys. } {\bf #1} (19#2) #3}

\end{document}